\documentclass[aip,jcp,reprint]{revtex4-1}

\usepackage{graphicx}
\usepackage{bm}
\newcommand{\subscript}[1]{\ensuremath{_{\textrm{\footnotesize{#1}}}}}
\newcommand{\ket}[1]{\left| #1 \right\rangle}

\newcommand{\bra}[1]{\left\langle #1 \right|}

\begin{document}

\title{Approaching exact hyperpolarizabilities via sum-over-states Monte Carlo configuration interaction}

\author{J. P. Coe}
\author{M. J. Paterson}%
 \email{M.J.Paterson@hw.ac.uk}
\affiliation{ 
Institute of Chemical Sciences, School of Engineering and Physical Sciences, Heriot-Watt University, Edinburgh, EH14 4AS, UK.
}%

\date{\today}

\begin{abstract}
We propose using sum-over-states calculations with the compact wavefunctions of Monte Carlo configuration interaction to approach accurate values for higher-order dipole properties up to second hyperpolarizabilities in a controlled way.  We apply the approach to small systems that can generally be compared with full configuration interaction (FCI) results. We consider hydrogen fluoride with a 6-31g basis and then look at results, including frequency dependent properties, in an aug-cc-pVDZ basis. We extend one calculation beyond FCI by using an aug-cc-pVTZ basis. The properties of an H\subscript{4} molecule with multireference character are calculated in an aug-cc-pVDZ basis. We then investigate this method on a strongly multireference system with a larger FCI space by modelling the properties of carbon monoxide with a stretched geometry. The behavior of the approach with increasing basis size is considered by calculating results for the neon atom using aug-cc-pVDZ to aug-cc-pVQZ. We finally test if the unusual change in polarizability between the first two states of molecular oxygen can be reproduced by this method in a 6-31g basis. 
\end{abstract}

\maketitle

\section{Introduction}

It has been demonstrated\cite{MCCIdipoles}  that multipole moments for a range of small molecules could generally be calculated to sufficient accuracy using Monte Carlo configuration interaction (MCCI).\cite{mcciGreer98,mccicodeGreer}  These calculations only used a very small fraction of the full configuration interaction (FCI) space and could be implemented, without the need for chemical intuition, even if the system was considered to be multireference.  However the calculation of higher-order properties, such as the dipole second hyperpolarizability, when using derivatives to fourth order in the energy  with regards to the electric field is more challenging and is particularly difficult for a stochastic method.  

These properties are of interest in photochemistry as, e.g., the nonlinear optical properties of a molecule may be exploited for frequency doubling of coherent light and can allow a molecular switch to be created.\cite{CHEM:CHEM2464,NLOswitch} Such switches have also been put forward as tools to detect the presence of certain molecules (see, for example, Ref.~\onlinecite{DetectionNLOswitches}).

Sum-over-states (SOS) allows one to computationally approach a property's value in a controlled manner where convergence with respect to the number of states can be monitored. Furthermore the important states for a qualitative description could perhaps be identified and by modifying the molecule to alter these states a nonlinear property may be tuned. However SOS is hampered by the need to compute many excited states to sufficient accuracy. This means that for systems for which FCI is possible, an accurate SOS calculation may still be computationally intractable.  MCCI has been demonstrated to find  wavefunctions that use only a very small fraction of the FCI space but capture much of the FCI result. We therefore investigate if the much smaller configuration space needed for MCCI allows SOS calculations to be feasible and accurate for small systems.

We use state-averaged MCCI\cite{2013saMCCI} for a small number of excited states to produce a tractable set of configurations that aims to capture enough of the ground and excited state aspects of the FCI wavefunction.  We then investigate the use of these configurations in computationally viable sum-over-states calculations for dipole polarizabilities, hyperpolarizabilities and second hyperpolarizabilities. Results for a selection of small molecules, including those with wavefunctions deemed to be significantly multireference in character, are compared with full configuration interaction values from numerical derivatives and coupled cluster\cite{CCSD,CCSD(T)} results using response methods.\cite{DaltonResponse} 

We first investigate this approach for hydrogen fluoride using the 6-31g basis, before calculating properties, both static and frequency dependent, in an aug-cc-pVDZ basis. The second hyperpolarizability calculation is extended to an aug-cc-pVTZ basis. A linear chain of four hydrogens is then considered and we look at carbon monoxide with a stretched geometry in the 6-31g basis to test the approach on strongly multireference problems. The effect of basis size is investigated with calculations on the neon atom ranging from aug-cc-pVDZ to aug-cc-pVQZ. Finally we consider the change in polarizability when going from the ground-state triplet of molecular oxygen to the first singlet state in a 6-31g basis.

\section{Methods}

We initially consider a time-independent homogeneous electric field applied along the z axis and use atomic units.  Within the dipole approximation this adds a term $F\hat{z}$ to the Hamiltonian where we have used $F$ for the electric field strength.  A Taylor series may be used to write the energy as a function of the electric field strength
\begin{equation}
E(F)=E(0)+\sum_{k=1}^{\infty}\frac{1}{k!}\frac{\partial^{k} E}{\partial F^{k}}F^{k}.
\label{eq:Efield}
\end{equation}

The permanent dipole moment in the z-direction is then identified as the negative of the first derivative $\mu_{0,z}=-\frac{\partial E}{\partial F}$.  While the polarizability $\alpha_{zz}=-\frac{\partial^{2} E}{\partial F^{2}}$, first hyperpolarizability $\beta_{zzz}=-\frac{\partial^{3} E}{\partial F^{3}}$ and second hyperpolarizability $\gamma_{zzzz}=-\frac{\partial^{4} E}{\partial F^{4}}$ are calculated similarly. To approximate these quantities, numerical derivatives may be used.
One may also consider the field strength as a perturbation parameter where the energy for state $i$ is
\begin{equation}
E_{i,PT}(F)=E_{i}+FE_{i}^{(1)}+F^{2}E_{i}^{(2)}+\cdots
\end{equation}
Perturbation theory may be used to calculate the energy to a given order. This can be used to calculate the property of interest by taking into account the appropriate factor in Eq.~\ref{eq:Efield}. Such an approach is known as sum-over-states (SOS).  For example, using
\begin{equation}
E_{0}^{(1)}=\sum_{k\neq 0} \frac{|\bra{\Psi_{0}} \hat{z} \ket{\Psi_{k}}|^{2}}{E_{0}-E_{k}}
\end{equation}
allows the ground-state polarizability to be calculated as $\alpha_{zz}=-2E_{0}^{(1)}$.  Here the $\Psi_{k}$ are the solutions of the unperturbed Hamiltonian.

Orr and Ward \cite{OrrAndWard} used time-dependent perturbation theory to derive expressions for the dependence of properties on the frequencies of the applied field. For the polarizability this leads to

\begin{equation}
\alpha_{zz}(\omega)=-\sum_{k\neq 0} \left(\frac{|\bra{\Psi_{0}} \hat{z} \ket{\Psi_{k}}|^{2}}{E_{0}-E_{k}-\omega} +\frac{|\bra{\Psi_{0}} \hat{z} \ket{\Psi_{k}}|^{2}}{E_{0}-E_{k}+\omega}\right).
\end{equation}

We note that we focus on higher-order dipole properties in this work, but in theory the approach could be used for higher-order multipole properties.

MCCI\cite{mcciGreer98,mccicodeGreer} has been used to construct wavefunctions that comprise a very small fraction of the FCI space yet often capture much of the FCI result.\cite{MCCIGreer95,dissociationGreer,GreerMcciSpectra,MCCIpotentials,MCCIdipoles,2013saMCCI}  MCCI has been also recently been used for the modelling of tunnel junctions,\cite{GreerComplexMCCI} transition metal dimers\cite{MCCImetaldimers} and the dissociation energies of first row diatomics.\cite{MCCIfirstrowdiss} An MCCI calculation is essentially determined by the cutoff parameter (c\subscript{min}). In this work the procedure begins with a configuration state function (CSF) formed from the occupied Hartree-Fock molecular orbitals. The MCCI space is stochastically enlarged using configurations formed from single and double substitutions chosen so that the wavefunction remains within a given spatial symmetry. The MCCI wavefunction is then found by diagonalization and any new configurations with absolute coefficient $|c_{i}|$ less than c\subscript{min} are deleted and the process continues. Here we use the approach of state-averaged MCCI\cite{2013saMCCI} to allow the stable calculation of excited states so $c_{i}=\sum_{j}|c_{i,j}|.$ where $j$ ranges over the excited states of interest. This process is repeated and  every 10 iterations all configurations become at risk of removal.  The MCCI energy is variational as it is found by diagonalization of a Hamiltonian matrix constructed in a subset of the FCI configuration space.  We note that one benefit of using state-averaged MCCI is that the initial states are treated in a balanced way. In this work we limit the number of initial states to eight, but this could be extended to systematically improve the results if time and computing resources allow. 

 Convergence in the MCCI calculation is appraised using the approach of Ref.~\onlinecite{GreerMcciSpectra}, but for all states of interest in this work where we use $5\times 10^{-4}$ Hartree as the energy convergence criterion. The Davidson-Liu algorithm\cite{LiuReportNumerical} is used to diagonalize the Hamiltonian matrix when dealing with multiple states during the MCCI calculation and also to calculate the relatively large number of excited states from the converged MCCI configuration space for use in the SOS computation. In this work we are concerned with benchmark calculations so tend to increase the number of states in the SOS calculation until the result appears to have converged with respect to the FCI results. In practice, for systems beyond FCI, a convergence check could be introduced for the unknown property to determine when to end a calculation. 

We are not stating that the SOS framework is preferable to using response functions, and note that in Ref.~\onlinecite{1992ResponseAndSOS} multiconfigurational self consistent field results demonstrated that convergence was slow for SOS compared with the response results for formaldehyde.  Earlier work\cite{SOSformaldehyde} using configuration interaction wavefunctions also found that SOS results using energy-ordered states converge slowly. However SOS calculations are relatively straightforward to implement when using configuration interaction wavefunctions and offer the prospect of identifying, then possibly tuning, important states. This approach has not been combined with a stochastic method before to our knowledge and the possibility of the compact MCCI wavefunction allowing tractable and accurate SOS calculations is worth investigating.

The Hartree-Fock molecular orbitals and their integrals are calculated using Columbus.\cite{Columbus}  The FCI\cite{MolproFCI1,MolproFCI2} and coupled cluster energy results in an electric field used to calculate the numerical derivatives are found using MOLPRO,\cite{MOLPRO_brief2012} while Dalton\cite{Dalton2013,DALTONPROG2013} is used for coupled cluster unrelaxed response calculations of properties.  

 A numerical derivative, even for very accurate methods, can give poor results for too large a step size.  However spurious values may occur if the step size is so small that it is close to the level of precision of the method. Based on preliminary work we use a step size of $0.005$ in the field strength and the finite difference equations given in the Appendix when calculating numerical derivatives.  We acknowledge that the use of numerical derivatives means that these FCI results for properties are not necessarily exact for a given basis.

In Ref.~\onlinecite{MCCImetaldimers} an approach to quantify the multireference character for a given basis and set of molecular orbitals was put forward $MR=\sum_{i} |c_{i}|^{2} -|c_{i}|^{4}$.  We employ this quantity in this work and note that as the value approaches unity then the system is characterized as being more multireference.

We summarize the SOS MCCI procedure below:
\begin{enumerate}
\item{State-averaged MCCI is used to find a set of configurations to initially describe the first $s$ states.}
\item{The Hamiltonian matrix is then constructed in this configuration space and diagonalized to give $M$ states where $M>>s$.}
\item{The SOS property is calculated as a function of the number of the $M$ states included.}
\end{enumerate} 

\section{Results}

\subsection{HF}

The ground state of hydrogen fluoride (HF) at its equilibrium geometry is considered to be well-described by a single reference so calculation of its properties would be better suited to methods such as coupled cluster.  However it serves as a useful test for the SOS MCCI method due the availability of FCI results.  The molecule is oriented along the z axis with the hydrogen atom before the fluorine atom.

\subsubsection{6-31g}

We initially use a bond length of $R=0.91$ \r{A} and the 6-31g basis with one frozen molecular orbital. The FCI configuration space is around $11,000$ Slater determinants (SDs) when symmetry is included. This means that the calculations are not onerous and we can relatively easily investigate the effect of varying the number of initial states or the cutoff on the results.  We initially look at the ground-state singlet of $A_{1}$ symmetry and note that when using the $C_{2v}$ point group only excited singlet states of $A_{1}$ symmetry can have $\bra{\Psi_{0}} \hat{z} \ket{\Psi_{i}}\neq 0$ and therefore contribute to the SOS property calculations.

When using eight initial states and c\subscript{min}$=5\times10^{-4}$ then the ground state has a dominant configuration with coefficient $0.98$ and a value of $0.21$ for its multireference character when using the Hartree-Fock molecular orbitals in the 6-31g basis. This supports the idea that this system may be well-described by methods based on a single-reference.

\begin{figure}[ht]\centering
\includegraphics[width=.45\textwidth]{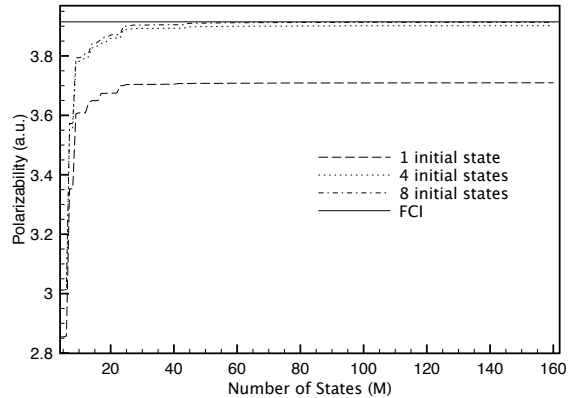}
\caption{HF SOS MCCI polarizability ($\alpha_{zz}$) results for c\subscript{min}$=5\times10^{-4}$ and various initial states (s) compared with the FCI numerical derivative result when using the 6-31g basis.}\label{fig:HF631gPolarise}
\end{figure}

In Fig.~\ref{fig:HF631gPolarise} we look at three initial MCCI calculations where one, four or eight states are considered in the production of the MCCI wavefunctions. States for use in the SOS calculation are then found by diagonalizing the Hamiltonian matrix constructed in the configuration space of the MCCI wavefunction.  The SOS polarizability is then plotted against the number of these states included. We see in Fig.~\ref{fig:HF631gPolarise} that the SOS MCCI polarizability appears to have converged when around 60 states are included. Essentially the FCI result is achieved on the scale of the graph when using 8 states for the MCCI wavefunction, while with 4 initial states the converged value is only a little less accurate. With 1 initial state the converged value is a little low.

\begin{figure}[ht]\centering
\includegraphics[width=.45\textwidth]{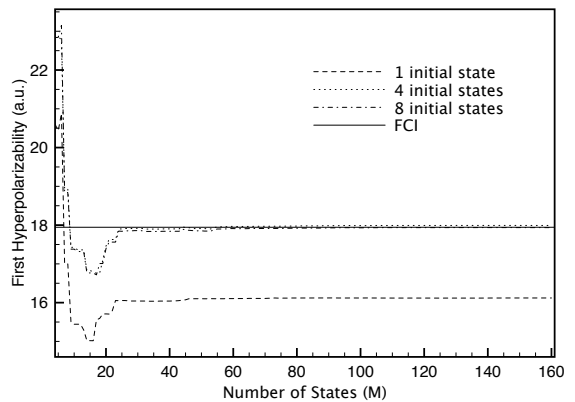}
\caption{Hydrogen fluoride SOS MCCI first hyperpolarizability ($\beta_{zzz}$) results for c\subscript{min}$=5\times10^{-4}$ and various initial states (s) compared with the FCI numerical derivative result when using the 6-31g basis.}\label{fig:HF631g1stHyper}
\end{figure}

A similar result is seen for the first hyperpolarizability (Fig.~\ref{fig:HF631g1stHyper}) where by 60 states in the SOS calculation convergence appears to have been reached. With 8 initial states the final result is almost indistinguishable from the FCI value in the graph and the value is only a little higher when using 4 initial states. 

\begin{figure}[ht]\centering
\includegraphics[width=.45\textwidth]{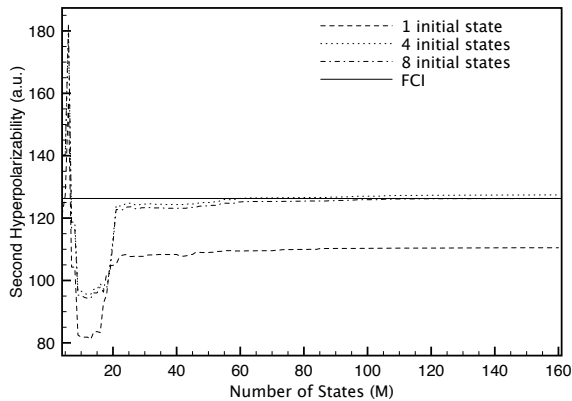}
\caption{Hydrogen fluoride  SOS MCCI second hyperpolarizability ($\gamma_{zzzz}$) results for c\subscript{min}$=5\times10^{-4}$ and various initial states (s) compared with the FCI numerical derivative result when using the 6-31g basis.}\label{fig:HF631g2ndHyper}
\end{figure}

The second hyperpolarizability seems to be more challenging in that convergence to approximately the FCI value for 8 initial states is not seen in Fig.~\ref{fig:HF631g2ndHyper} until around 120 states are included in MCCI SOS.

\begin{figure}[ht]\centering
\includegraphics[width=.45\textwidth]{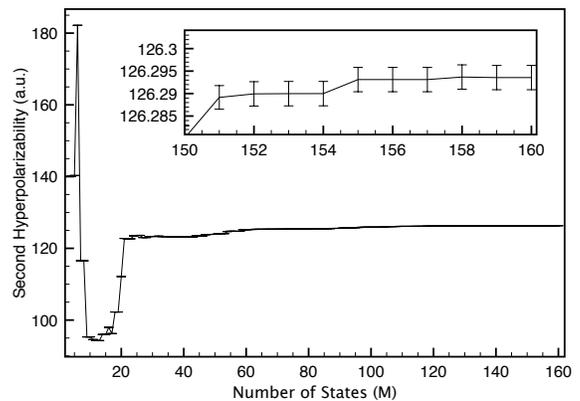}
\caption{Hydrogen fluoride mean SOS MCCI second hyperpolarizability ($\gamma_{zzzz}$) results from ten MCCI calculations with different random number seeds for c\subscript{min}$=5\times10^{-4}$ with eight initial states. The standard error is displayed by the error bars. Inset: Enlarged view of the results when the last ten states are included in the SOS calculation.}\label{fig:HF_631g_2ndHyper10runsErrors}
\end{figure}

The coefficients in the MCCI wavefunction are found by a deterministic diagonalization and one could fix the random number seed in MCCI so that the converged energy will always be the same for a given number of processors. However as MCCI uses a stochastic process to find the configuration space in which the wavefunction is calculated it is of interest to observe the effect of changing the random number seed on the SOS MCCI results.  We consider the second hyperpolarizability and run MCCI ten times with different random number seeds.  We see in Fig.~\ref{fig:HF_631g_2ndHyper10runsErrors} that for this system, with a reasonable cutoff, the standard errors in the mean SOS results are very small on the scale of the graph and an enlarged view of the inclusion the last ten states reveals that the standard errors are less than the small change in the property when going from 150 to 160 states.  This fits in with earlier work where in Ref.~\onlinecite{MCCIpotentials} it was observed that the error between FCI and MCCI results for potential curves of hydrogen-fluoride was also not affected much by changing the random number seed.

If we use numerical derivatives of the MCCI energy when varying the field strength then the polarizability result is fairly accurate at $3.93$ (a.u.), but the first hyperpolarizability is too large at $69$ (a.u.) and the second hyperpolarizability highlights the difficulty of higher order numerical derivatives as it is much too large and of the wrong sign at $-36676$ (a.u.).  We note that if analytic derivatives can be incorporated into MCCI then this may allow higher derivatives to be computed accurately from a single MCCI calculation.

In Table \ref{tbl:ResultsHF631gGround} we see that with 8 initial states then for the three properties essentially FCI results have been recovered when 160 states are used for the SOS calculation using around $13\%$ of the configurations in the relatively small FCI space. Using 4 initial states the results are a little less accurate but require fewer configurations while with 1 initial state the fraction recovered is always less than $95\%$ and this decreases with higher order properties.

\begin{table}[h]
\centering
\caption{Fraction of FCI (three decimal places) recovered by SOS MCCI with 160 total states for the $A_{1}$ ground state of hydrogen fluoride with the 6-31g basis. SDs are used for FCI while CSFs are used for MCCI.} \label{tbl:ResultsHF631gGround}
\begin{tabular*}{8.5cm}{@{\extracolsep{\fill}}lcccc}
\hline
\hline
 & $\alpha_{zz}$ & $\beta_{zzz}$ & $\gamma_{zzzz}$ & Configurations  \\
\hline
1 initial state   & $0.948$  & $0.899$ & $0.875$ & $0.049$  \\
4 initial states   & $0.997$  & $1.003$  & $1.009$ & $0.112$ \\
8 initial states   & $1.000$  & $0.999$  & $1.000$ & $0.130$ \\
\hline
\hline
\end{tabular*}
\end{table}

\begin{figure}[ht]\centering
\includegraphics[width=.45\textwidth]{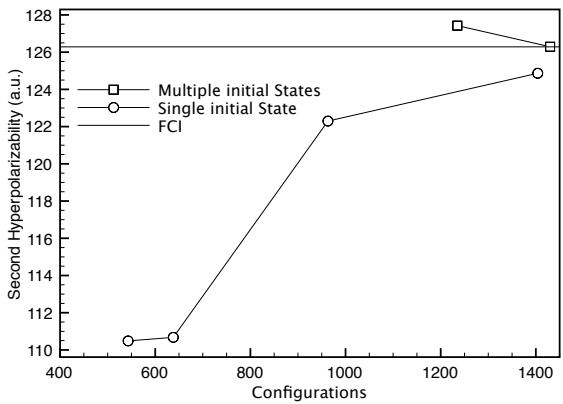}
\caption{Comparison of SOS MCCI second hyperpolarizability ($\gamma_{zzzz}$) results for hydrogen fluoride  with the 6-31g basis and 160 final states when multiple initial states (s) are used with c\subscript{min}$=5\times10^{-4}$ with a single initial state when c\subscript{min} is varied versus the number of configurations in the MCCI wavefunction.}\label{fig:HF631g2ndHyperVcsfs}
\end{figure}

We briefly consider if there is a benefit of using MCCI with a number of states rather than a small cutoff and one state.  We have seen that the properties are more accurate when more states are considered in the initial MCCI calculation, but this comes with an increase in the number of configurations that need to be considered. The improvement in accuracy could therefore be perhaps solely attributed to more configurations being included thereby giving a result closer to that of FCI.  We hypothesize, however, that there are important configurations for excited states that are required to give a good description of higher-order properties in an SOS calculation that will not be included until a very low cut-off is used in the ground state. In Fig.~\ref{fig:HF631g2ndHyperVcsfs} we plot the second hyperpolarizability against the number of configurations when using more initial states compared with lowering the cutoff for one initial state. The results suggest that to reach the FCI value would require more configurations when lowering c\subscript{min} than when increasing the number of initial states. It therefore does appear more efficient in this case to consider excited states in the initial calculation although the difference is not as great as we expected. We continue by carrying out initial calculations of SA-MCCI with eight states.

\begin{figure}[ht]\centering
\includegraphics[width=.45\textwidth]{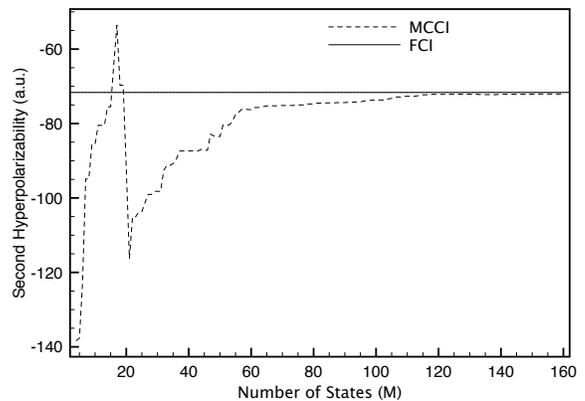}
\caption{Hydrogen fluoride  first $A_{1}$ singlet excited state second hyperpolarizability ($\gamma_{zzzz}$) from SOS MCCI with c\subscript{min}$=5\times10^{-4}$ and eight initial states versus number of states compared with the FCI result when using the 6-31g basis.}\label{fig:HF_631g_2ndHyperExcited}
\end{figure}

We now also calculate results for the first excited singlet state of $A_{1}$ symmetry. With the SOS approach this does not require a further MCCI calculation.  When using eight initial states then the first excited state has the same value as the ground state ($0.21$) for its multireference character so also appears to not require multireference methods. The results for the most challenging quantity ($\gamma_{zzzz}$) are displayed in Fig.~\ref{fig:HF_631g_2ndHyperExcited}.  While the ground-state second hyperpolarizability had reached $97.5\%$ of the FCI value by 40 states, the first excited state second hyperpolarizability has only reached $82.0\%$ by 40 states. This suggests that the excited state may be a little more challenging for the SOS MCCI approach. Table \ref{tbl:ResultsHF631gExcited} shows that when 160 states are included then the properties are very close to that of FCI but a little less accurate than those of the ground-state when 8 initial states are employed.

\begin{table}[h]
\centering
\caption{Fraction of FCI (three decimal places) recovered by SOS MCCI with 160 total states for the first excited $A_{1}$ state of hydrogen fluoride with the 6-31g basis. SDs are used for FCI while CSFs are used for MCCI.} \label{tbl:ResultsHF631gExcited}
\begin{tabular*}{8.5cm}{@{\extracolsep{\fill}}lcccc}
\hline
\hline
 & $\alpha_{zz}$ & $\beta_{zzz}$ & $\gamma_{zzzz}$ & Configurations  \\
\hline
8 initial states   & $0.997$  & $0.999$  & $0.994$ & $0.130$ \\
\hline
\hline
\end{tabular*}
\end{table}

\subsubsection{aug-cc-pVDZ}

We now consider the aug-cc-pVDZ basis with one frozen molecular orbital and a bond length of $1.7328795$ Bohr.  This enables us to compare our SOS MCCI results with FCI calculations in Ref.~\onlinecite{HFandNeandBHhyperFCI}.  We note our finite difference results for the FCI polarizability and hyperpolarizability when using a step size of $0.005$ are in agreement to two decimal places with those of Ref.~\onlinecite{HFandNeandBHhyperFCI}.  The FCI configuration space is now of the order of $10^{8}$ Slater determinants when symmetry is included.  The multireference character for the MCCI results in this basis is $0.30$ for  c\subscript{min}$=5\times10^{-4}$ 
 suggesting that methods based on a single-reference would still be expected to perform well in this case.

\begin{figure}[ht]\centering
\includegraphics[width=.45\textwidth]{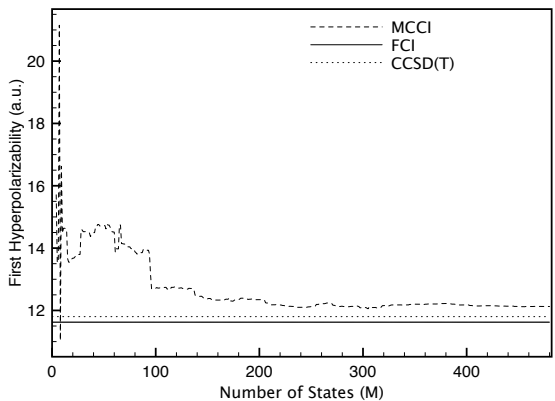}
\caption{Hydrogen fluoride first hyperpolarizability from SOS MCCI with 8 initial states and c\subscript{min}$=5\times10^{-4}$ compared with numerical derivative results from FCI and CCSD(T) plotted against the total number of states using the aug-cc-pVDZ basis.}\label{fig:HFaugvdzHyper}
\end{figure}

For the first hyperpolarizability (Fig.~\ref{fig:HFaugvdzHyper}) we see that convergence of the SOS property is now much slower in the larger basis: more than 200 states are required for the SOS MCCI result to appear that convergence is being approached. The result is slightly less close to the FCI value than when using the smaller basis and we see that, as would be expected, the numerical derivative of CCSD(T) results is more accurate than SOS MCCI. We note that the CCSD response result at $12.14$ (a.u.)  is similar to the SOS MCCI value.

\begin{figure}[ht]\centering
\includegraphics[width=.45\textwidth]{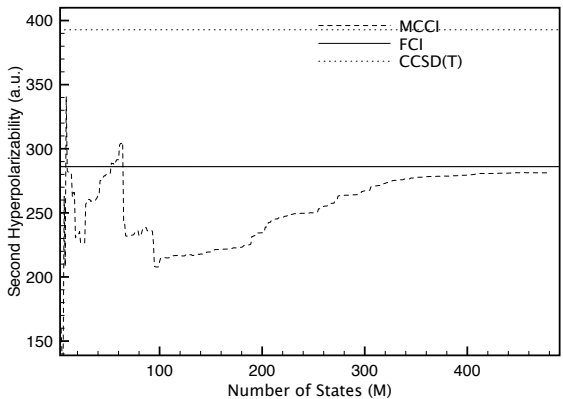}
\caption{Hydrogen fluoride second hyperpolarizability from SOS MCCI with 8 initial states and c\subscript{min}$=5\times10^{-4}$ compared with numerical derivative results from FCI and CCSD(T) plotted against the total number of states using the aug-cc-pVDZ basis.}\label{fig:HFaugvdz2ndHyper}
\end{figure}

In Fig.~\ref{fig:HFaugvdz2ndHyper} we see that convergence is slower for the second hyperpolarizability, but the SOS MCCI result appears close to FCI when more than 400 states are employed in total. The CCSD(T) property is less accurate the SOS MCCI in this case.  This may demonstrate one of the problems of using numerical derivatives.  The CCSD response results are similar to CCSD numerical derivative calculations except for the second hyperpolarizability which is negative. Our other preliminary results suggested that this may be a sign convention in the version of Dalton used.\cite{Dalton2013,DALTONPROG2013} Hence in this work we take the opposite sign of the CCSD response second hyperpolarizability and note that at $305$ (a.u.)  it represents $107\%$ of the FCI result compared with $98\%$ for SOS MCCI and is closer than the numerical derivatives.

For the highest order property considered we see that now more than 400 states are necessary to seemingly reach convergence. As we appear to have converged but still have a noticeable, although small, discrepancy with the FCI results then this suggests that a lower cutoff rather than more states in the SOS calculation may be necessary to improve the results. We investigate lowering the cutoff to c\subscript{min}$=2\times10^{-4}$ and find that the multireference character of the MCCI wavefunction only changes slightly to $0.32$. We also now consider frequency dependent results for the polarizability and first hyperpolarizability. For the frequency dependent first hyperpolarizability for a field in the z-direction we use the expression of Ref.~\onlinecite{HFandNeandBHhyperFCI}:

\begin{eqnarray}
\nonumber \beta_{zzz}(-2\omega;\omega,\omega)=2\sum_{i,j\neq0} \frac{\bra{\Psi_{0}} \hat{z} \ket{\Psi_{i}}\bra{\Psi_{i}} \bar{z} \ket{\Psi_{j}} \bra{\Psi_{j}} \hat{z} \ket{\Psi_{0}} }{(E_{i}-E_{0}-2\omega)(E_{0}-E_{j}+\omega)}\\
\nonumber +\frac{\bra{\Psi_{0}} \hat{z} \ket{\Psi_{i}}\bra{\Psi_{i}} \bar{z} \ket{\Psi_{j}} \bra{\Psi_{j}} \hat{z} \ket{\Psi_{0}} }{(E_{i}-E_{0}+\omega)(E_{0}-E_{j}-2\omega)}\\
\nonumber +\frac{\bra{\Psi_{0}} \hat{z} \ket{\Psi_{i}}\bra{\Psi_{i}} \bar{z} \ket{\Psi_{j}} \bra{\Psi_{j}} \hat{z} \ket{\Psi_{0}} }{(E_{i}-E_{0}+\omega)(E_{0}-E_{j}+\omega)}
\end{eqnarray}
where $\bar{z}=\hat{z}-\bra{\Psi_{0}} \hat{z} \ket{\Psi_{0}}$.

\begin{figure}[ht]\centering
\includegraphics[width=.45\textwidth]{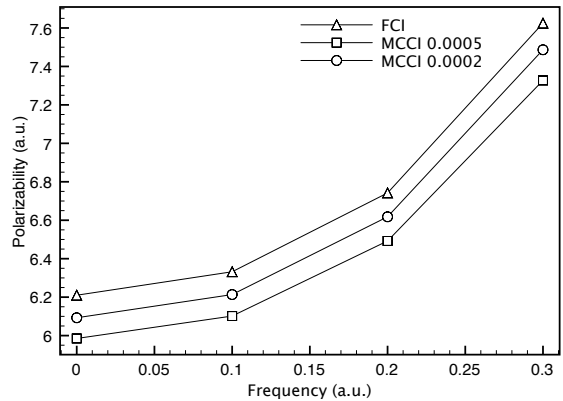}
\caption{Hydrogen fluoride polarizability from SOS MCCI with 8 initial states and $480$ total states compared with FCI results from Ref.~\onlinecite{HFandNeandBHhyperFCI} plotted against frequency $\omega$ using the aug-cc-pVDZ basis.}\label{fig:HF_augvdz_PolarVw}
\end{figure}

In Fig.~\ref{fig:HF_augvdz_PolarVw} we see that the shape of the FCI frequency dependency curve for the polarizability is reproduced by SOS MCCI. The results underestimate the FCI values a little but are improved by lowering the cutoff from c\subscript{min}$=5\times10^{-4}$ to c\subscript{min}$=2\times10^{-4}$.

\begin{figure}[ht]\centering
\includegraphics[width=.45\textwidth]{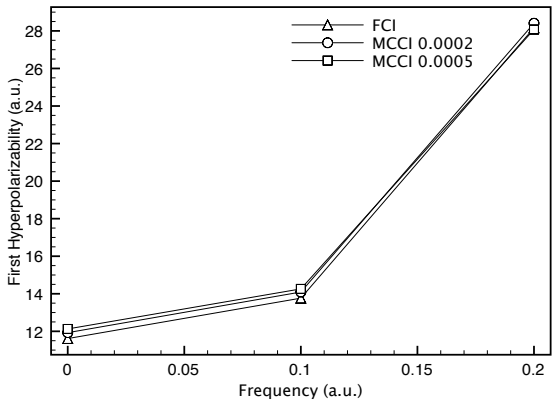}
\caption{Hydrogen fluoride hyperpolarizability from SOS MCCI with 8 initial states and $480$ total states compared with FCI results from Ref.~\onlinecite{HFandNeandBHhyperFCI} plotted against frequency $\omega$ using the aug-cc-pVDZ basis.}\label{fig:HF_augvdz_hyperPolarVw}
\end{figure}

The first hyperpolarizability frequency dependency curve (Fig.~\ref{fig:HF_augvdz_hyperPolarVw}) is also generally recovered by SOS MCCI.  The difference between FCI and SOS MCCI appears less on the scale of the graph compared with the polarizability and it is not clear if lowering the cutoff to $=2\times10^{-4}$ has improved the results at $\omega=0.2$ when 480 states are used.

By comparing Tables \ref{tbl:HFaugccpvdz} and \ref{tbl:HFaugccpvdzSmallercutoff}  we see that lowering the cutoff generally improves the values so that they are all within $3\%$ of FCI but the static second hyperpolarizability and the first hyper polarizability with $\omega=0.2$ are a little less accurate at the lower cut-off. This is only a small difference which may be due to the non-variational nature of these properties or it could be that more than 480 states need to be considered at the lower cut-off to reach a sufficiently converged value.

\begin{table}[h]
\centering
\caption{Fraction to three decimal places of FCI results recovered by SOS MCCI with 480 total states and c\subscript{min}$=5\times10^{-4}$  for the first excited $A_{1}$ state of hydrogen fluoride with the aug-cc-pVDZ basis for a range of frequencies ($\omega$). FCI results for $\omega \neq 0$ are from Ref.~\onlinecite{HFandNeandBHhyperFCI}} \label{tbl:HFaugccpvdz}
\begin{tabular*}{8.5cm}{@{\extracolsep{\fill}}lccc}
\hline
\hline
 $\omega$ & $\alpha_{zz}$ & $\beta_{zzz}$ & $\gamma_{zzzz}$  \\
\hline
$0$   & $0.964$   & $1.043$  & $0.983$   \\
$0.1$ & $0.964$  & $1.036$ &  -  \\
$0.2$ &  $0.963$ & $0.997$  & -  \\
$0.3$ &  $0.961$ & -       & -  \\
\hline
\hline
\end{tabular*}
\end{table}

\begin{table}[h]
\centering
\caption{Fraction to three decimal places of FCI results recovered by SOS MCCI with 480 total states and c\subscript{min}$=2\times10^{-4}$  for the first excited $A_{1}$ state of hydrogen fluoride with the aug-cc-pVDZ basis for a range of frequencies ($\omega$). FCI results for $\omega \neq 0$ are from Ref.~\onlinecite{HFandNeandBHhyperFCI}} \label{tbl:HFaugccpvdzSmallercutoff}
\begin{tabular*}{8.5cm}{@{\extracolsep{\fill}}lccc}
\hline
\hline
 $\omega$ & $\alpha_{zz}$ & $\beta_{zzz}$ & $\gamma_{zzzz}$  \\
\hline
$0$   & $0.981$   & $1.026$ & $0.970$   \\
$0.1$ & $0.981$  & $1.024$  &  - \\
$0.2$ & $0.982$  & $1.009$  &  - \\
$0.3$ & $0.982$  &  -      &  -  \\
\hline
\hline
\end{tabular*}
\end{table}

 We note that at the larger cutoff the number of CSFs used in MCCI is 10664 which represents around $4\times 10^{-3}\%$  of the FCI Slater determinant space while for the smaller cutoff, 26386 CSFs were used which is about $0.01\%$ of the FCI space.  

If we extend the work to the aug-cc-pVTZ basis then the MCCI results at c\subscript{min}$=5\times10^{-4}$  used $18102$ configurations compared with the FCI space of $~10^{11}$ Slater determinants.  We find that 480 states gives an SOS MCCI second hyperpolarizability of $\gamma_{zzzz}=307.62$ (a.u.).  This is in reasonable agreement with the numerical derivative CCSD(T) results ($\gamma_{zzzz}=323.82$ (a.u.) ) and the CCSD response value of $\gamma_{zzzz}=330.90$ (a.u.). 

The SOS MCCI values could perhaps be improved by lowering the cutoff further and considering more states. However we do not pursue this as we have shown that MCCI can give properties using SOS values to high accuracy for HF and other methods would be more appropriate for very accurate calculations on this single-reference system.

\subsection{H\subscript{4}}

We now compare the SOS MCCI results with finite-field FCI calculations for H\subscript{4} in an aug-cc-pVDZ basis with no frozen molecular orbitals. We consider two hydrogens with bond length $r_{2}$ each bonded to another hydrogen at a distance of $r_{1}$ in a linear chain oriented along the z-axis. Here $r_{1}=1.9$ \r{A} and  $r_{2}=1.7$ \r{A}.

When calculating the FCI energy and using numerical derivatives we find the negative of the fourth derivative to be six times larger than second hyperpolarizability in Ref.~\onlinecite{H4secondhyper} due to different definitions in terms of the Taylor expansion of the dipole (negative derivative of Eq.~\ref{eq:Efield} with respect of F). We therefore translate the result of Ref.~\onlinecite{H4secondhyper} to the definition we have used for $\gamma$.

When using the $D_{2h}$ point group, then for a ground state of symmetry $A_{g}$ only excited states of $A_{g}$ and $B_{1u}$ symmetries may have non-zero values for the integrals $\bra{\Psi_{0}} \hat{z} \ket{\Psi_{k}}$. The eight initial states consist of 4 states for each of the two symmetries.  We then use the resulting set of configurations to calculate 20 states of each symmetry. These states are ordered by their energy.

\begin{figure}[ht]\centering
\includegraphics[width=.45\textwidth]{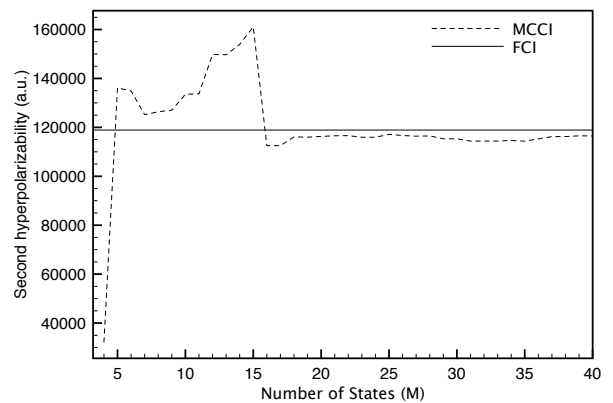}
\caption{Second hyperpolarizability $\gamma_{zzzz}$ of H\subscript{4} from SOS MCCI with eight initial states and c\subscript{min}$=5\times10^{-4}$ versus number of states compared with numerical derivative FCI results\cite{H4secondhyper} for the aug-cc-pVDZ basis.}\label{fig:H4hyper}
\end{figure}

Ref.~\onlinecite{H4secondhyper} finds that for an SOS FCI calculation then 36 states are required to converge to the numerical derivative result for the second hyperpolarizability.  There it was noted that 18 FCI states were required to give $95\%$ of the numerical derivative result.  In Fig.~\ref{fig:H4hyper} we see that the SOS MCCI value becomes close to the FCI result when 20 states are included and then the result seems to oscillate a little as the number of states heads towards 40.  With 18 MCCI states we recover most of the FCI result ($97.6\%$) and the MCCI result is only a little closer by the time that 40 states are used ($98.9\%$) for this non-variational quantity. MCCI used 2522 CSFs used on average for the two symmetries compared with around $5\times 10^{4}$ SDs for FCI. 

We calculate the SOS MCCI polarizability at c\subscript{min}$=5\times10^{-4}$ as 51.73 using 40 states while the numerical derivative value for FCI is $52.93$.  By 20 states $96.2\%$ of the FCI value has been recovered compared with $97.7\%$ by 40 states. 

The ground-state multireference character is found to be $0.84$ for the MCCI result when using the aug-cc-pVDZ basis. We note that we recover a similar fraction of the FCI results as for HF with an aug-cc-pVDZ basis despite the FCI space being smaller for H\subscript{4}. We suggest that the strongly multireference character makes this system more challenging despite the relatively small FCI configuration space.  We next consider a system with both multireference character and a reasonably large FCI space.

\subsection{CO}

Carbon monoxide at a stretched geometry of $4$ Bohr would be expected to require multireference methods to model the system accurately.  In Ref.~\onlinecite{MCCIdipoles} it was found that MCCI could give the FCI dipole with an error of $1.70\%$ when using a cc-pVDZ basis while the CCSD result was around four times too large. 

We investigate if SOS MCCI can calculate higher order properties sufficiently close to FCI when using a 6-31g basis set and two frozen orbitals. With c\subscript{min}$=5\times10^{-4}$  and eight initial states we find that the ground-state of this system should indeed be considered strongly multireference in this basis with Hartree-Fock molecular orbitals due to its multireference character of $0.91$. 

\begin{figure}[ht]\centering
\includegraphics[width=.45\textwidth]{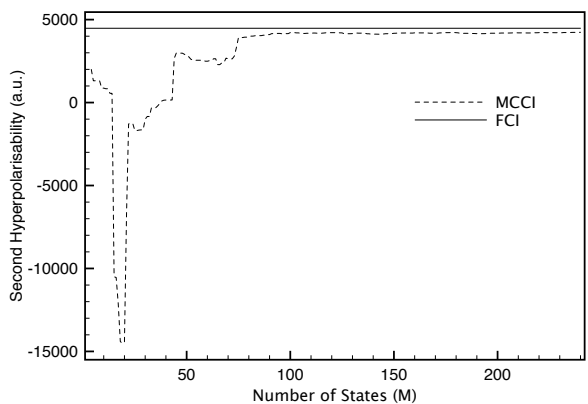}
\caption{CO second hyperpolarizability from SOS MCCI with 8 initial states and c\subscript{min}$=5\times10^{-4}$ compared with numerical derivative results plotted against the total number of states using the 6-31g basis.}\label{fig:CO_631g_2ndHyper}
\end{figure}

Fig.~\ref{fig:CO_631g_2ndHyper} shows that the SOS MCCI second hyperpolarizability is reasonably close to the FCI result on the scale of the graph. The SOS MCCI results appear to have converged when around 100 states are included.  Interestingly the value is much too large and has the wrong sign when only around 20 states are considered.

The MCCI results used $6507$ CSFs compared with around $5\times10^{6}$ Slater determinants for the FCI space when using symmetry.  In table \ref{tbl:CO631g} the SOS MCCI results are seen to be close to FCI but are the least accurate of the systems and bases considered at this point. We attribute this to the strongly multireference nature of this example coupled with the reasonably large FCI space.  CCSD response results are surprisingly close to FCI for $\alpha_{zz}$ and $\beta_{zzz}$  where they surpass the accuracy of MCCI for the former.  However,  we note that, in line with previous work\cite{MCCIdipoles} with the cc-pVDZ basis, the CCSD dipole is three times too large compared with FCI.  Furthermore the result for $\gamma_{zzzz}$ has the wrong sign compared with FCI and its magnitude is also strongly in error.

\begin{table}[h]
\centering
\caption{Fraction of FCI (two decimal places) recovered by SOS MCCI results with 8 initial state, c\subscript{min}$=5\times10^{-4}$ then 240 total states and CCSD unrelaxed response for the ground state of CO with $R=4$ Bohr and the 6-31g basis.} \label{tbl:CO631g}
\begin{tabular*}{8.5cm}{@{\extracolsep{\fill}}lccc}
\hline
\hline
 & $\alpha_{zz}$ & $\beta_{zzz}$ & $\gamma_{zzzz}$ \\
\hline
MCCI   & $0.94$  & $1.13$  &  $0.95$ \\
CCSD   & $0.99$  & $ 1.13$ & -$4.40$       \\
\hline
\hline
\end{tabular*}
\end{table}

\section{Neon atom}

The experimentally measurable second hyperpolarizability of the neon atom has been calculated to high accuracy using coupled cluster response methods with large bases, see for example Refs.~\onlinecite{1998NeSecondHyper,2004NeSecondHyper}.  We now use the neon atom to investigate the effect of basis size on the convergence and accuracy of the SOS MCCI second hyperpolarizability $\gamma_{zzzz}$ compared with CCSD response calculations. For the ground state of $A_{g}$ symmetry in $D_{2h}$ then only states of $A_{g}$ and $B_{1u}$ contribute to $\gamma_{zzzz}$.  We compute an equal number of each symmetry for SOS MCCI and use c\subscript{min}$=5\times10^{-4}$ with one frozen molecular orbital.

\begin{figure}[ht]\centering
\includegraphics[width=.45\textwidth]{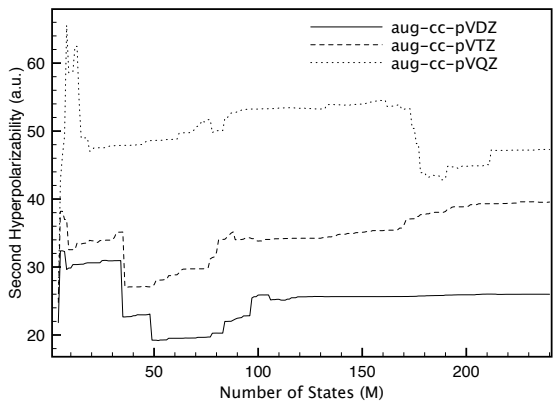}
\caption{Neon second polarizability from SOS MCCI with 8 initial states and c\subscript{min}$=5\times10^{-4}$ plotted against the total number of states for aug-cc-pVXZ basis sets.}\label{fig:Ne_2ndhyper}
\end{figure}

In Fig.~\ref{fig:Ne_2ndhyper} we see that by around 150 states the aug-cc-pVDZ property appears to have essentially converged. It is not clear if the results in aug-cc-pVTZ are close to their converged value after 200 state are included and the noticeable drop in the aug-cc-pVQZ result when around 175 states are included suggests that more than 240 states may be necessary in the largest basis set considered.
We note that the CCSD response result in the aug-cc-pVDZ basis represents $100.6\%$ of the numerical derivative FCI results. The larger bases are beyond our current FCI capabilities so we now compare the SOS MCCI results with those of CCSD response calculations.  We see that the fraction of the second hyperpolarizability recovered by SOS MCCI does not decrease smoothly with increasing basis size (table~\ref{tbl:NeResults}): the most accurate result is in the aug-cc-pVTZ basis set. This may be fortuitous in that the fully converged SOS MCCI value could perhaps be less accurate. The least accurate result is in the aug-cc-pVQZ basis where only around $10^{-6}\%$ of the FCI space is used. This may have been anticipated if we are including less of the FCI space for a system whose correlation would be expected to be categorized as dynamic.

\begin{table}[h]
\centering
\caption{Fraction of CCSD response $\gamma_{zzzz}$ (three decimal places) calculated by SOS MCCI with 240 total states and c\subscript{min}$=5\times10^{-4}$ for the ground state of the neon atom and fraction of the FCI configurations used by MCCI. SDs are used for FCI while CSFs are used for MCCI.} \label{tbl:NeResults}
\begin{tabular*}{8.5cm}{@{\extracolsep{\fill}}lcc}
\hline
\hline
 Basis & $\gamma_{zzzz}$ & Configurations  \\
\hline
aug-cc-pVDZ   & $0.878$  & $\sim10^{-4}$ \\
aug-cc-pVTZ   & $0.941$  & $\sim 10^{-6}$ \\
aug-cc-pVQZ   & $0.813$  & $\sim 10^{-8}$ \\
\hline
\hline
\end{tabular*}
\end{table}

We quantify this idea by calculating the multireference nature of the MCCI results. In order of increasing basis size we find the values to be $0.24$, $0.25$ and $0.24$.  This shows that the correlation would be classified as dynamic and the system should be well modelled by an approach based on a single reference.  Hence SOS MCCI would not be the method of choice for this system and goes someway to explaining why only around $80\%$ of the CCSD response result is captured in the largest basis when using a fixed cut-off. However the SOS MCCI results do appear to be heading in the correct direction with regards to basis size (Fig.~\ref{fig:Ne_2ndhyper}) as Ref.~\onlinecite{2004NeSecondHyper} finds the best theoretical estimate to be about $108$ (a.u.) using very large basis sets.

\section{O\subscript{2}}

One unconventional feature of molecular oxygen is that the ground-state polarizability has been calculated to be greater than the first excited state.\cite{Poulsen98,Minaev2007}  We now investigate if SOS MCCI with c\subscript{min}$=5\times10^{-4}$ can reproduce the difference in polarizabilities compared with numerical derivative FCI results in the 6-31g basis with two frozen molecular orbitals. The first excited state is a singlet while the ground-state is a triplet which limits the application of coupled cluster response methods in this instance. For the triplet of $B_{1g}$ symmetry in $D_{2h}$ only states of $B_{1g}$ and $A_{u}$ symmetry can be combined to give nonzero integrals in the calculation of $\alpha_{zz}$ while for the $A_{g}$ singlet, states of $A_{g}$ and $B_{1u}$ are required. We equally partition the eight initial states between the two required symmetries.

\begin{figure}[ht]\centering
\includegraphics[width=.45\textwidth]{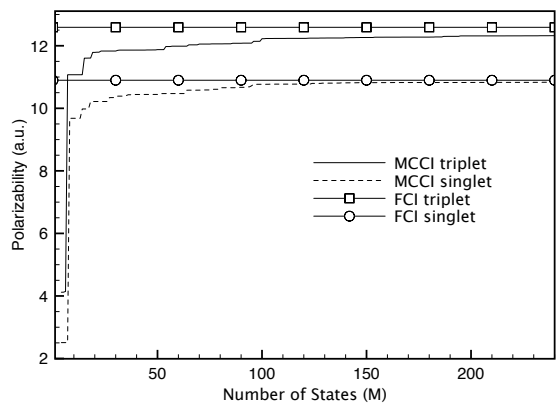}
\caption{O\subscript{2} polarizability for the first triplet and singlet states from SOS MCCI with 8 initial states and c\subscript{min}$=5\times10^{-4}$ compared with FCI numerical derivative results plotted against the total number of states using the 6-31g basis.}\label{fig:O2_631g_Polarise}
\end{figure}

In Fig.~\ref{fig:O2_631g_Polarise} we see that SOC MCCI with the 6-31g basis can qualitatively capture the curious behavior of the polarizability in the first two states of molecular oxygen. The singlet result appear to be almost that of FCI when more than 150 states are included.  There is a more noticeable gap between SOS MCCI and FCI for the triplet results even when more than 200 states are considered.

For the singlet polarizability we recover $99.4\%$ of the FCI result, while $97.9\%$ is recovered for the triplet state. We quantify the multireference nature of the singlet as $0.74$ and the triplet as $0.48$ when using CSFs. It appears that the singlet is multireference but not strongly so and the triplet would perhaps not be classified as multireference. This may suggest that more of the correlation could be classified as dynamic for the triplet so, at the reasonable cutoff employed, may be neglected by MCCI and could be the reason that SOS performs slightly less well.  The MCCI calculation used on average 5500 CSFs. The FCI space consisted of approximately eight million SDs while the MCCI results used around $0.07\%$ of the FCI space.

\section{Summary}

We have put forward the idea that sum-over-states (SOS) may be used with the compact wavefunctions resulting from Monte Carlo configuration interaction (MCCI) to enable tractable calculations of properties up to the dipole second hyperpolarizability.  We saw that with hydrogen fluoride when using the 6-31g basis that essentially full configuration interaction (FCI) results could be recovered using around $13\%$ of the fairly small FCI space when $\sim 120$ states are used for SOS calculations. Results suggested that computing a reasonable number of states in the initial state-averaged MCCI calculation may be preferable to considering only one state and lowering the cutoff. We then used an aug-cc-pVDZ basis and also considered frequency dependent results. We found that more than 400 states were necessary to converge the SOS values when around $4\times10^{-3}\%$ of the approximately $10^{8}$ configurations in the FCI space were used. Property values were slightly less accurate in this larger basis but within around $4\%$ and were generally improved by lowering the cutoff.  We also briefly considered the aug-cc-pVTZ basis to push the problem beyond what is currently feasible with our current FCI calculations. In this case, reasonable agreement for the second hyperpolarizability between SOS MCCI and coupled cluster approaches was observed.
  
A multireference system with an FCI space of $\sim 5\times 10^{4}$ was then considered: H\subscript{4} with a stretched geometry. Here we recovered more than $97\%$ of the FCI property values by using 40 states. 

We next looked at carbon monoxide with a geometry of $4$ Bohr and a 6-31g basis. This system was found to be strongly multireference and has a relatively large FCI space of $5\times 10^{6}$ Slater determinants. CCSD response results were accurate for some properties in this system, but had severe difficulties with others while the MCCI SOS results were all within $13\%$  of the FCI values.  

To investigate the effect of increasing the basis size we considered the second hyperpolarizability of the neon atom. There we found that convergence appeared to take longer as the basis size was increased. The largest basis (aug-cc-pVQZ) gave the least accurate result when compared with CCSD response for this system which is expected to be very well-described by methods based on a single reference. The trend in the property appeared to be in the right direction with increasing basis compared with the literature value when very large basis sets were employed.

Finally we demonstrated that the unconventional lowering of the polarizability in the first excited state of molecular oxygen could be qualitatively reproduced by SOS MCCI in a 6-31g basis. The results were at least $97.9\%$ of the FCI values despite using only a very small fraction of the FCI space.

In contrast to numerical differentiation, we demonstrated that these calculations can approach the full configuration interaction value in a controlled way.  We saw that SOS MCCI fared well for all considered properties of a stretched CO molecule. This result for a system with a large amount of static correlation leads us to suggest that SOS MCCI has potential for applications to larger systems that are challenging for methods built around a single reference and are computationally intractable for FCI.  A convergence check for the SOS property could be used for future work but is not considered here as the goal is comparison with FCI.  By using the SOS framework there is the possibility for understanding which states are important for a quantity and how this quantity may be manipulated by modifying these states through changing aspects of the molecule. Natural transition geminals\cite{NatGeminals} could perhaps be used to characterize the important states in terms of single and double excitations from the ground-state.  However we note that the number of states necessary for convergence seemed to increase with basis size for a given system and this could eventually limit the generalization of SOS MCCI. Finally, we note that the approach is not limited to higher-order dipole properties and, e.g, quadrupole polarizabilities could be calculated as up to octupole moments were demonstrated to be able to be modelled by MCCI in Ref.~\onlinecite{MCCIdipoles}.

\acknowledgements{We thank the European Research Council (ERC) for funding under the European Union's Seventh Framework Programme (FP7/2007-2013)/ERC Grant No. 258990.}

\section{Appendix}
We use the following finite difference formulae to compute the numerical derivatives
\begin{equation}
f'(x)\approx\frac{f(x+h)-f(x-h)}{2h},
\end{equation}

\begin{equation}
f''(x)\approx\frac{f(x+h)-2f(x)+f(x-h)}{h^{2}},
\end{equation}

\begin{eqnarray}
\nonumber f'''(x)\approx \frac{1}{2h^{3}} \big( f(x+2h)-2f(x+h)\\
+2f(x-h)-f(x-2h) \big),
\end{eqnarray}

\begin{eqnarray}
\nonumber f''''(x)\approx \frac{1}{h^{4}}\big(f(x+2h)-4f(x+h)+6f(x)\\
-4f(x-h)+f(x-2h)\big).
\end{eqnarray}

\providecommand{\noopsort}[1]{}\providecommand{\singleletter}[1]{#1}%

\end{document}